\documentclass[eqsecnum,10pt,aps,prd]{revtex4}
\begin{document}
\title{Reality of linear and angular momentum expectation values in
bound states}
\author{Utpal Roy}
\email{utpalroy@prl.res.in}
\affiliation{Physical Research Laboratory, Ahmedabad 380009, India}
\author{Suranjana Ghosh}
\email{sanjana@prl.res.in}
\affiliation{Physical Research Laboratory, Ahmedabad 380009, India}
\author{T. Shreecharan}
\email{shreet@prl.res.in}
\affiliation{Physical Research Laboratory, Ahmedabad 380009, India}
\author{Kaushik Bhattacharya}
\email{kaushik@nucleares.unam.mx}
\affiliation{Insituto de Ciencias Nucleares, Universidad
Nacional Aut\'{o}noma de Mexico, Circuito Exterior, C.U., A. Postal 70-543, C. Postal 04510, Mexico DF, Mexico}
%%%%%%%%%%%%%%%%%%%%%%%%%%%%%%%%%%%%%%%%%%%%%%%%%%%%%
\begin{abstract}
\begin{center}
{\small ABSTRACT}
\end{center}
In quantum mechanics textbooks the momentum operator is defined in the
Cartesian coordinates and rarely the form of the momentum operator in
spherical polar coordinates is discussed. Consequently one always
generalizes the Cartesian prescription to other coordinates and falls
in a trap. In this work we introduce the difficulties one faces when
the question of the momentum operator in spherical polar coordinate
comes. We have tried to point out most of the elementary quantum
mechanical results, related to the momentum operator, which has
coordinate dependence. We explicitly calculate the momentum
expectation values in various bound states and show that the
expectation value really turns out to be zero, a consequence of the
fact that the momentum expectation value is real.  We comment briefly
on the status of the angular variables in quantum mechanics and the
problems related in interpreting them as dynamical variables.  At the
end, we calculate the Heisenberg's equation of motion for the radial
component of the momentum for the Hydrogen atom.
\end{abstract}
\maketitle
%%%%%%%%%%%%%%%
\section{Introduction}
%%%%%%%%%%
Quantum mechanics is a treasure house of peculiar and interesting
things. Elementary textbooks of quantum mechanics \cite{element1,
element2, element3} generally start with the postulates which are
required to define the nature of the dynamical variables in the
theory and their commutation relations. The choice of the
dynamical variables is not that clear, as the coordinates in
Cartesian system are all elevated to the status of operators where
as time remains a parameter. More over in spherical polar
coordinates only the radial component can be represented as an
operator while the angles still remain as a problem. The
difficulty of giving different status to the spatial coordinates
and time is bypassed in quantum field theories where all the
coordinates and time become parameters of the theory. But the
problem with angles still remain a puzzle which requires to be
understood in future.

When we start to learn quantum mechanics, most of the time we
begin with elementary calculations relating to the particle in a
one dimensional infinite well, particle in a finite potential
well, linear harmonic oscillator and so on. The main aim of these
calculations is to solve the Schr\"{o}dinger equation in the
specific cases and find out the bound state energies and the
energy eigenfunctions in coordinate space representation. While
solving these problems we overlook the subtleties of other quantum
mechanical objects as the definition of the momentum operator in
various coordinates, the reality of its expectation value, etc..
In the last one or two decades there has been a number of studies
regarding the self-adjointness of various operators \cite{gjg}.
The aim of these studies has been to analyze the self-adjointness
of various operators like momentum, Hamiltonian etc. and find out
whether these operators are really self-adjoint in some interval
of space where the theory is defined, if not then can there be any
mathematical method by which we can make these operators to be
self-adjoint in the specified intervals ?

In the present work we deal with a much elementary concept in
quantum mechanics related to the reality of the expectation values
of the momentum operator, be it linear or angular. We do not
analyze the self-adjointness of the operators which requires
different mathematical techniques. To test the self-adjointness of
an operator we have to see whether the operator is symmetric in a
specific spatial interval and the functional domain of the
operator and its adjoint are the same. In the article we always
keep in touch with the recent findings from modern research about
self-adjoint extensions but loosely we assume that the operators
which we are dealing with are Hermitian. If something is in
contrary we point it out in the main text. A considerable portion
of our article deals with the analysis of the fact that the
expectation value of the momentum operator in various bound states
are zero, a result which most of the textbooks only quote but
never show. In the simpler cases the result can be shown by one or
two lines of calculation, but in nontrivial potentials as the
Morse potential, the Coulomb potential the result is established
by using various properties of the special functions such as the
associated Legendre and the associated Laguerre.

The presentation of various materials in our article is done in the
following way. Next section deals with the definition of the momentum
operator and its properties. Section \ref{intri} deals with the
intricacies of the definition of the momentum operator in spherical
polar coordinates and the problems we face when we try to mechanically
implement the quantization condition, which is invariably always
written in Cartesian coordinates in most of the textbooks on quantum
mechanics. In section \ref{momexpv} we explicitly calculate the
momentum expectation values in various potentials and show that in
bound states we always get the expectation value of the linear
momentum to be zero. Section \ref{ether} gives a brief discussion on
the Ehrenfest theorem when we are using it to find out the time
derivative of the expectation value of the radial component of
momentum in the case of the Hydrogen atom. We end with the concluding
section which summarizes the findings in our article.

Before going into the main discussion we would like to mention
about the convention. We have deliberately put a hat over various
symbols to show that they are operators in quantum mechanics. Some
times this convention becomes tricky when we are dealing with
angular variables as there the status of these variables is in
question. The other symbols have their conventional meaning. As we
are always using the coordinate representation  sometimes we
may omit the hat over the position operator as in this
representation the position operator and its eigenvalues can be
trivially interchanged.
%%%%%%%%%%%%%%%%%%%
\section{Definition of the momentum operator and the reality of its
expectation value}
\label{defmom}
%%%%%%%%%%%%%%%
From the Poisson bracket formalism of classical mechanics we can infer:
\begin{eqnarray}
[\hat{x}_i , \hat{p}_j] = i\hbar\,\delta_{i\,j}\,,
\label{bcomm}
\end{eqnarray}
where $\delta_{i\,j}=1$ when $i=j$ and zero for all other cases, and
$i,j=1,2,3$. In the above equation $\hat{x}_i$ is the position
operator and $\hat{p}_j$ is the linear momentum operator in Cartesian
coordinates. From the above equation we can also find the form of the
momentum operator in position representation, which is:
\begin{eqnarray}
\hat{p_{i}}=-i\hbar \frac{\partial}{\partial x_{i}}\,.
\label{momdef}
\end{eqnarray}
It is interesting to note that the above expression of the momentum
operator also gives us the form of the generator of translations.
This is because of the property:
\begin{eqnarray}
[\hat{p}_x , F(\hat{x})]= -i\hbar \frac{d F(\hat{x})}{dx}\,,
\label{gen}
\end{eqnarray}
where $F(\hat{x})$ is an arbitrary well defined function of $\hat{x}$. The
above equation ensures that the momentum operator generates
translations along the $x$ direction.

Particularly in one-dimension the expression of the momentum operator
becomes $\hat{p}_x=-i\hbar \frac{\partial}{\partial x}$. We know that
the expectation value of the momentum operator must be real. If we
focus on one-dimensional systems to start with, where the system is
specified by the wave-function $\psi(x,t)$, the expectation value of
any operator $\hat{O}$ is defined by:
\begin{eqnarray}
\langle \hat{O} \rangle \equiv \int^\infty_{-\infty} \psi^*(x,t)\,\hat{O}
\,\psi(x,t)\,dx\,,
\label{exptv}
\end{eqnarray}
where $\psi^{*}$ signifies complex conjugation of $\psi$ and the
extent of the system is taken as $-\infty < x < \infty$. From the
above equation we can write,
\begin{eqnarray}
\langle \hat{O} \rangle^* = \int^\infty_{-\infty}
\psi(x,t)\,\hat{O}^* \,\psi(x,t)^*\,dx\,. \label{cexptv}
\end{eqnarray}
If $\langle \hat{O} \rangle = \langle \hat{O} \rangle^*$ then the
condition of the reality of the expectation value becomes:
\begin{eqnarray}
\int^\infty_{-\infty} \psi^*(x,t)\,\hat{O}
\,\psi(x,t)\,dx = \int^\infty_{-\infty} \psi(x,t)\,\hat{O}^*
\,\psi(x,t)^*\,dx\,.
\label{hermcond}
\end{eqnarray}
For a three-dimensional system the above condition becomes,
\begin{eqnarray}
\int^\infty_{-\infty} \psi^*({\bf x},t)\,\hat{O}
\,\psi({\bf x},t)\,d^3 x = \int^\infty_{-\infty} \psi({\bf x},t)\,\hat{O}^*
\,\psi({\bf x},t)^*\,d^3 x\,.
\label{jhermcond}
\end{eqnarray}
Now if we take the specific case of the momentum operator in
one-dimension we can explicitly show that its expectation value is
real if the extent of the system is infinite and the wave-function
vanishes at infinity. The proof is as follows. The expectation value
of the momentum operator is:
\begin{eqnarray}
\int^\infty_{-\infty} \psi^*(x,t)\,\hat{p}_x\,\psi(x,t)\,dx
&=& -i\hbar\int^\infty_{-\infty} \psi^*(x,t)\,
\frac{\partial\psi(x,t)}{\partial x}\,dx\nonumber\\
&=& -i\hbar \left[\left.\psi^*(x,t)\psi(x,t)\right|^\infty_{-\infty} -
\int^\infty_{-\infty}\psi(x,t)\frac{\partial\psi(x,t)^*}{\partial x}\,dx
\right]\,,
\label{momexp}
\end{eqnarray}
If the wave functions vanish at infinity then the first term on the
second line on the right-hand side of the above equation drops and we
have,
\begin{eqnarray}
\int^\infty_{-\infty} \psi^*(x,t)\,\hat{p}_x\,\psi(x,t)\,dx
&=& -i\hbar\int^\infty_{-\infty} \psi^*(x,t)\,
\frac{\partial\psi(x,t)}{\partial x}\,dx\nonumber\\
&=& i\hbar \int^\infty_{-\infty}\psi(x,t)\frac{\partial\psi(x,t)^*}{\partial x}\,dx\,,\nonumber\\
&=&\int^\infty_{-\infty} \psi(x,t)\,\hat{p}_x^*\,\psi(x,t)^*\,dx\,.
\label{hermomexp}
\end{eqnarray}
A similar proof holds for the three-dimensional case where it is
assumed that the wave-function vanishes at the boundary surface at
infinity.
%%%%%%%%%%%%%%%%%%%%%%%%%%%%%%%%%%%%%%%%%%
\section{The expectation value of the momentum
operator in Cartesian and spherical polar coordinates}
\label{intri}
%%%%%%%%%%%%%%%%%
In non-relativistic version of quantum mechanics we know that if we
have a particle of mass $m$ which is present in a time-independent
potential we can separate the Schr\"{o}dinger equation:
\begin{eqnarray}
i\hbar\frac{\partial\psi({\bf x},t)}{\partial t}=\left[-\frac{\hbar^2}{2m}
{\bf \nabla}^2 + V({\bf x})\right]\psi({\bf x},t)\,,
\label{scheq}
\end{eqnarray}
into two equations, one is the time-dependent one which gives the trivial
solution $e^{-\frac{iEt}{\hbar}}$ where $E$ is the total energy of the
particle, and the other equation is the time-independent Schr\"{o}dinger equation:
\begin{eqnarray}
{\bf \nabla}^2 u({\bf x})+\frac{2m}{\hbar^2}(E-V({\bf x}))u({\bf x})=0\,,
\label{tise}
\end{eqnarray}
where $u({\bf x})$ is the solution of the time-independent Schr\"{o}dinger
equation and the complete solution of the Eq.~(\ref{scheq}) is:
\begin{eqnarray}
\psi({\bf x},t)=u({\bf x}) e^{-\frac{iEt}{\hbar}}\,.
\label{ssol}
\end{eqnarray}
In the case of the free-particle, where $V({\bf x})=0$, we have
$u({\bf x},t)=e^{i{\bf k}\cdot{\bf x}}$ where
$E=\frac{k^2\hbar^2}{2m}$ and $k=|{\bf k}|$. The free-particle
solution is an eigenfunction of the momentum operator with eigen
value $\hbar k$. Although if we try to find out the expectation
value of the momentum operator as is done in the last section we
will be in trouble as these wave-functions do not vanish at
infinity, a typical property of free-particle solutions. But this
problem is not related to the Hermiticity property of the momentum
operator, it is related with the de-localized nature of the
free-particle solution.

In physics many times we require to solve a problem using curvilinear
coordinate systems. The choice of our coordinate system depends upon
the specific symmetry which we have at hand. Suppose we are working in
spherical polar coordinates and the solution of Eq.~(\ref{tise}) can
be separated into well behaved functions of $r$, $\theta$ and
$\phi$ as,
\begin{eqnarray}
u({\bf x}) = u(r,\theta,\phi)= R(r)\,\Theta(\theta)\,\Phi(\phi)\,.
\label{sepvar}
\end{eqnarray}
If we try to follow the proof of the Hermiticity of the linear
momenta components, as done in the last section, in spherical
polar coordinates, then we should write:
\begin{eqnarray}
\langle \hat{\bf p} \rangle &=& -i\hbar \int_{\tau}
u^{*}(r,\theta,\phi)\nabla u(r,\theta,\phi)\,d\tau\,,\nonumber\\
&=& -i\hbar \int R^*(r)\Theta^*(\theta)\Phi^*(\phi)\ \left[{\bf
e}_r\frac{\partial}{\partial r} + \frac{{\bf e}_\theta}{r}
\frac{\partial}{\partial \theta} + \frac{{\bf e}_\phi}{r
\sin\theta} \frac{\partial}{\partial
\phi}\right]R(r)\Theta(\theta)\Phi(\phi)
r^2 dr d\Omega\,,\nonumber\\
\label{pspc}
\end{eqnarray}
where in the above equation ${\bf e}_r$, ${\bf e}_\theta$, ${\bf
e}_\phi$ respectively are the unit vectors along $r$, $\theta$ and
$\phi$, and $d\Omega=\sin\theta \,d\theta \,d\phi$. $\tau$ is the
volume over which we integrate the expression in the above
equation. From the last equation we can write:
\begin{eqnarray}
\langle \hat{p}_r \rangle = -i\hbar \left[ \int_\Omega |\Theta(\theta)|^2
|\Phi(\phi)|^2 d\Omega \right]
\int^{\infty}_0 r^2 R^*(r)\frac{d R(r)}{d r}\,dr\,,
\end{eqnarray}
As $\Theta(\theta)$ and $\Phi(\phi)$ are normalized, the
integration: $\int_\Omega |\Theta(\theta)|^2|\Phi(\phi)|^2 d\Omega
=1$ and we can proceed as in Eq.~(\ref{hermomexp}) as:
\begin{eqnarray}
\langle \hat{p}_r \rangle &=& -i\hbar
\int^{\infty}_0 r^2 R^*(r)\frac{d R(r)}{d r}\,dr\,,\nonumber\\
&=&-i\hbar \left[ \left.r^2 R^*(r) R(r)\right|^\infty_0  -
\int^\infty_0 \left(2r R^*(r) + r^2 \frac{d R^*(r)}{dr}\right)R(r)\,dr
\right]\,.
\label{rherm}
\end{eqnarray}
If $R(r)$ vanishes at infinity then the above equation reduces to,
\begin{eqnarray}
\langle \hat{p}_r \rangle &=& \left[ i\hbar \int^\infty_0 r^2 R(r)
\frac{d R^*(r)}{dr}\,dr \right] + 2 i\hbar \int^\infty_0 r |R(r)|^2 \,dr\,,
\nonumber\\
&=& \langle \hat{p}_r \rangle^* + 2 i\hbar \int^\infty_0 r |R(r)|^2 \,dr\,.
\label{nonherm}
\end{eqnarray}
The above equation implies that $\langle\hat{p}_r\rangle$ is not real
in spherical polar coordinates. The solution of the above problem lies
in redefining $\hat{p}_r$ as is evident from Eq.~(\ref{nonherm}), and
it was given by Dirac \cite{dirac, sieg}. The redefined linear
momentum operator along $r$ can be:
\begin{eqnarray}
\hat{p}_r \equiv -i\hbar\left(\frac{\partial}{\partial r} +
\frac{1}{r}\right)= -i\hbar \frac{1}{r}\frac{\partial}{\partial r}\,r\,.
\label{remom}
\end{eqnarray}
This definition of the $\hat{p}_r$ is suitable because in this form it
satisfies the commutation relation as given in Eq.~(\ref{bcomm}) where
now the operator conjugate to $\hat{r}$ is $\hat{p}_r$. The form of
$\hat{p}_r$ in Eq.~(\ref{remom}) shows that for any arbitrary function
of $r$ as $F(r)$ we must still have Eq.~(\ref{gen}) satisfied. This
implies that the modified form of $\hat{p}_r$ is still a generator of
translations along the $r$ direction. Up to this point we were
following what was said by Dirac regarding the status of the radial
momentum operator. Still everything is not that smooth with the
redefined operator as we can see that it turns out to be singular
around $r=0$, more over, although the radial momentum acts like
a translation generator along $r$ but near $r=0$ it cannot generate a
translation towards the left as the interval ends there.

In this regard we can state that the issue of the reality of the
radial component of the momentum in spherical polar coordinates is a
topic of modern research in theoretical physics \cite{paz1, paz2}. It
has been shown that the operator $-i\hbar \frac{\partial}{\partial r}$
is not Hermitian and more over it can be shown \cite{gjg} that such an
operator cannot be self-adjoint in the interval $[0 , \infty]$. In
some recent work \cite{paz2} the author claims that there can be an
unitary operator which connects $-i\hbar\frac{\partial}{\partial r}$
to $-i\hbar \frac{1}{r}\frac{\partial}{\partial r}\,r$, and as the
former operator does not have a self-adjoint extension in the
semi-infinite interval so the latter is also not self-adjoint in the
same interval.

If we further try to find out whether $\langle\hat{p}_\theta\rangle$
and $\langle\hat{p}_\phi\rangle$ are real, then we will face
difficulties. Working out naively if we claim that $\hat{p}_\phi=
\frac{1}{rsin\theta} \frac{\partial}{\partial \phi}$ as suggested by
the $\phi$ component of Eq.~(\ref{pspc}) we will notice that
$\phi\,\hat{p}_\phi$ does not have the dimension of action. This means
$\hat{p}_\phi$ or $\hat{p}_\theta$ is not conjugate to $\phi$ or
$\theta$. This is a direct representation of the special coordinate
dependence of the quantization condition. Only in Cartesian
coordinates the variables conjugate to $x$, $y$ and $z$ are $p_x$,
$p_y$ and $p_z$. Taking the clue from classical mechanics we
know the proper dynamical variables conjugate to $\hat{\phi}$ and
$\hat{\theta}$ are the angular momentum operators, namely
$\hat{L}_\theta$ and $\hat{L}_\phi$. In general $\hat{L}_\phi$ is
given by:
\begin{eqnarray}
\hat{L}_\phi=-i\hbar\frac{\partial}{\partial \phi}\,,
\label{lphi}
\end{eqnarray}
which can be shown to posses real expectation values by following a
similar proof as is done in Eq.~(\ref{momexp}) and
Eq.~(\ref{hermomexp}), if we assume $\Phi(0)=\Phi(2\pi)$.  In this form
it is tempting to say that we can have a relation of the form,
\begin{eqnarray}
[\hat{\phi}, \hat{L}_\phi]=i\hbar\,,
\end{eqnarray}
which looks algebraically correct. But the difficulty in writing
such an equation is in the interpretation of $\hat{\phi}$ which
has been elevated from an angular variable to a dynamical
operator. In spherical polar coordinates both $\theta$ and $\phi$
are compact variables and consequently have their own subtleties.
Much work is being done in trying to understand the status of
angular variables and phases \cite{phases1, phases2}, in this work
we only present one example showing the difficulty of accepting
$\hat{\phi}$ as an operator.

From the solution of the time-independent Schr\"{o}dinger equation for an
isotropic potential we will always have:
\begin{eqnarray}
\Phi(\phi)=\frac{1}{\sqrt{2\pi}}\,e^{iM \phi}\,,
\label{phieqn}
\end{eqnarray}
where $M=0,\pm 1,\pm 2, \cdot, \cdot$. Now if $\hat{\phi}$ is an
operator we can find its expectation value, and it turns out to be:
\begin{eqnarray}
\langle \hat{\phi} \rangle &=& \frac{1}{2\pi}
\int^{2\pi}_0 \phi e^{iM\phi} e^{-iM\phi}\,d\phi\,,\nonumber\\
&=& \pi \,,
\end{eqnarray}
and the expectation value of $\hat{\phi}^2$ is:
\begin{eqnarray}
\langle \hat{\phi}^2 \rangle &=& \frac{1}{2\pi}
\int^{2\pi}_0 \phi^2 e^{iM\phi} e^{-iM\phi}\,d\phi\,,\nonumber\\
&=& \frac{4}{3}\pi^2 \,.
\end{eqnarray}
Consequently $\Delta \phi = \sqrt{\langle \hat{\phi}^2 \rangle -
\langle \hat{\phi} \rangle^2} = \frac{\pi}{\sqrt{3}}$. Similarly
calculating $\hat{L}_\phi$ we get:
\begin{eqnarray}
\langle \hat{L}_\phi \rangle &=& \frac{M\hbar}{2\pi}
\int^{2\pi}_0 e^{-iM\phi} e^{iM\phi}\,d\phi\,,\nonumber\\
&=& M\hbar \,,
\end{eqnarray}
as expected, and $\langle \hat{L}^2_\phi \rangle = M^2\hbar^2$. This
implies $\Delta L_\phi =\sqrt{\langle \hat{L}^2_\phi \rangle - \langle
\hat{L}_\phi \rangle^2} = 0$. So we can immediately see that the
Heisenberg uncertainty relation between $\hat{\phi}$ and
$\hat{L}_\phi$, $\Delta \phi \Delta L_\phi \geq \hbar/2$
breaks down. This fact makes life difficult and we have no means to
eradicate this problem.

Taking the clue from the $\phi$ part we can propose that
$\hat{L}_\theta$ is also of the form $-i\hbar \frac{\partial}{\partial
\theta}$. With this definition of $\hat{L}_\theta$ let us try to prove
its Hermitian nature as done in Eq.~(\ref{rherm}). Taking $R(r)$ and
$\Phi(\phi)$ in Eq.~(\ref{sepvar}) separately normalized, we can write:
\begin{eqnarray}
\langle \hat{L}_\theta \rangle &=& -i\hbar \int^{\pi}_0 \Theta^*(\theta)
\frac{d\Theta(\theta)}{d\theta}\,\sin\theta d\theta\,\nonumber\\
&=& -i\hbar\left[\left.\sin\theta \,\Theta^*(\theta)
\Theta(\theta)\right|^\pi_0 -  \int^{\pi}_0 \left(
\cos\theta \,\Theta^*(\theta) + \sin\theta
\frac{d\Theta^*(\theta)}{d\theta}\right)\Theta(\theta)\,d\theta\right]\,,
\nonumber\\
&=& \left[i\hbar\int^{\pi}_0 \sin\theta\,\Theta(\theta)
\frac{d\Theta^*(\theta)}{d\theta}\,d\theta \right]
+ i\hbar \int^{\pi}_0 \cos\theta\,|\Theta(\theta)|^2\,d\theta\,,\nonumber\\
&=& \langle \hat{L}_\theta \rangle^* + i\hbar \int^{\pi}_0 \cos\theta\,|\Theta(\theta)|^2\,d\theta\,.
\end{eqnarray}
The above equation shows that $\langle\hat{L}_\theta\rangle$ is not
real.  The rest is similar to the analysis following
Eq.~(\ref{nonherm}) where now we have to redefine the angular momentum
operator conjugate to $\theta$ as \cite{essen}:
\begin{eqnarray}
\hat{L}_\theta \equiv -i\hbar\left(\frac{\partial}{\partial \theta}
+ \frac12 \cot\theta \right)\,.
\end{eqnarray}
Unlike the $\phi$ case, $\Theta(\theta)$ are not eigenfunctions of
$\hat{L}_\theta$. But the difficulties of establishing $\theta$ as an
operator still persists and in general $\theta$ is not taken to be a
dynamical operator in quantum mechanics.

It is known that both $\theta$ and $\phi$ are compact variables,
i.e. they have a finite extent. But there is a difference between
them. In spherical polar coordinates the range of $\phi$ and $\theta$
are not the same, $0 \leq \phi < 2\pi$ and $0 \leq \theta \leq
\pi$. This difference can have physical effects. As $\phi$ runs over
the whole angular range so the wave-function corresponding to it
$\Phi(\phi)$ is periodic in nature while due to the range of $\theta$,
$\Theta(\theta)$ need not be periodic. Consequently there can be a net
angular momentum along the $\phi$ direction while there cannot be any
net angular momentum along $\theta$ direction. And this can be easily
shown to be true.  As the time-independent Schr\"{o}dinger equation for an
isotropic potential yields $\Phi(\phi)$ as given in Eq.~(\ref{phieqn})
similarly it is known that in such a potential the form of
$\Theta(\theta)$ is given by:
\begin{eqnarray}
\Theta(\theta)=N_\theta\,P^L_M(\cos\theta)\,,
\label{thetaeqn}
\end{eqnarray}
where $N_\theta$ is a normalization constant depending on $L$, $M$
and $P^L_M(\cos\theta)$ is the associated Legendre function, which
is real. In the above equation $L$ and $M$ are integers where
$L=0,1,2,3, \cdot, \cdot$ and $M=0, \pm 1, \pm 2, \pm 3, \cdot,
\cdot$. The quantum number $M$ appearing in Eq.~(\ref{phieqn}) and
in Eq.~(\ref{thetaeqn}) are the same. This becomes evident when we
solve the time-independent Schr\"{o}dinger equation in spherical
polar coordinates by the method of separation of variables. A
requirement of the solution is $-L \leq M \leq L$. Now we can
calculate the expectation value of $\hat{L}_\theta$ using the
above wave-function and it is:
\begin{eqnarray}
\langle \hat{L}_\theta \rangle &=& -i\hbar N^2_\theta\int^\pi_0
P^L_M(\cos\theta) \left(\frac{d P^L_M(\cos\theta)}{d\theta} +
\frac12 \cot\theta P^L_M(\cos\theta)\right)
\sin\theta d\theta\,\nonumber\\
&=& -i\hbar N^2_\theta\left[\int^\pi_0
P^L_M(\cos\theta)\frac{d P^L_M(\cos\theta)}{d\theta}
\sin\theta d\theta
+ \frac12 \int^\pi_0 P^L_M(\cos\theta) P^L_M(\cos\theta) \cos\theta \,d\theta
\right]\,.\nonumber\\
\end{eqnarray}
 To evaluate the integrals on the right hand side of the
above equation we can take $x=\cos\theta$ and then the expectation
value becomes:
\begin{eqnarray}
\langle \hat{L}_\theta \rangle &=& -i\hbar N^2_{\theta}\left[
\int^{-1}_1 P^L_M(x) \frac{d P_L^M (x)}{dx}
(1-x^2)^{\frac12}\,dx\right.\nonumber\\
&-&\left.\frac{1}{2}\int^{-1}_{1} P^{L}_{M(x)}
P^{L}_{M(x)}\frac{x}{\sqrt{1-x^{2}}} dx \right]\,. \label{postel}
\end{eqnarray}
The second term in the right hand side of the above equation
vanishes as the integrand is an odd function in the integration
range. For the first integral we use the following recurrence
relation \cite{grad}:
\begin{equation}
(x^2 - 1) \frac{d P^L_M (x)}{dx} =  M x P^L_M (x)
- (L+ M) P^L_{M-1}(x)\,,
\label{asslegen}
\end{equation}
the last integral can be written as,
\begin{eqnarray}
\langle \hat{L}_\theta \rangle &=& i\hbar N^2_\theta\left[M \int^{-1}_1
x (1 - x^2)^{-\frac12} P^L_M(x)  P^L_M (x)\,dx \right.\nonumber\\
&-&\left. (L+M) \int^{-1}_1
(1 - x^2)^{-\frac12} P^L_M (x) P^L_{M-1}(x)\,dx \right]\,.
\end{eqnarray}
As,
\begin{eqnarray}
P^L_M(x)= (-1)^{L+M}P^L_M(-x)\,,
\label{ppar}
\end{eqnarray}
 we can see immediately that both the integrands in the right hand
side of the above equation is odd and consequently $\langle
\hat{L}_\theta \rangle = 0$ as expected. A similar analysis gives
$\langle \hat{L}_\phi \rangle = M\hbar$. It must be noted that the form
of $\hat{L}_\theta$ still permits it to be the generator of rotations
along the $\theta$ direction.

As the motion along $\phi$ is closed so there can be a net flow of
angular momentum along that direction but because the motion along
$\theta$ is not so, a net momentum along $\theta$ direction will not
conserve probability and consequently for probability conservation we
must have expectation value of angular momentum along such a direction
to be zero. In elementary quantum mechanics text books it is often
loosely written that the solution of the time-independent
Schr\"{o}dinger equation is real when we are solving it for a real
potential. But this statement is not correct. The reality of the
solution also depends upon the coordinate system used. Specially for
compact periodic coordinates we can always have complex functions as
solutions without breaking any laws of physics.

Before leaving the discussion on angular variables in spherical polar
coordinates we want to point out one simple thing which is
interesting. In Cartesian coordinates when we deal with angular
momentum we know that:
\begin{eqnarray}
[\hat{L}_i, \hat{L}_j]=i\epsilon_{ijk}\,\hat{L}_k\,,
\end{eqnarray}
where $\hat{L}_i$ stands for $\hat{L}_x$, $\hat{L}_y$ or
$\hat{L}_z$. For this reason there cannot be any state which can be
labelled by the quantum numbers of any two of the above angular
momenta. But from the expressions of $\hat{L}_\phi$ and
$\hat{L}_\theta$ we see that,
\begin{eqnarray}
[\hat{L}_\phi, \hat{L}_\theta] = 0\,,
\end{eqnarray}
and consequently in spherical polar coordinates we can have
wave-function solutions of the Schr\"{o}dinger equation which are
simultaneous eigenfunctions of both $\hat{L}_\phi$ and
$\hat{L}_\theta$ as $P^L_M(\theta)$.

For real $V({\bf x})$, we expect the solution of the time-independent
Schr\"{o}dinger equation $u({\bf x})$ to be real, when we are solving
the problem in Cartesian coordinates. In all these cases the
expectation value of the linear momentum operators must vanish. The
reason is simple and can be understood in one-dimensional cases where
with real $u(x)$ we directly see that the integral
$\int^\infty_{-\infty} u^*(x)\,\frac{\partial u(x)}{\partial x}\, dx$
is real and so $\int^\infty_{-\infty} u^*(x)\,\hat{p}_x\,u(x)\,dx$
becomes imaginary as $\hat{p}_x$ contains $i$, as is evident from the
first line in Eq.~(\ref{hermomexp}). So if the expectation value of
the momentum operator has to be real then the only outcome can be that
for all those cases where we have a time-independent solution in a
bounded region of space, with a real potential and working in
Cartesian coordinates, the expectation value of the momentum operator
must vanish. The above statement is true in curvilinear coordinates
also, but in those cases the definition of the momentum operators have
to be modified. This fact becomes clear when we write the relationship
between the probability flux and the expectation value of the
momentum operator. The probability flux for a particle of mass $m$
is:
\begin{eqnarray}
{\bf j}({\bf x}, t) &=& -\left(\frac{i\hbar}{2m}\right)\left[
\psi^*({\bf x},t) \nabla \psi({\bf x},t) - (\nabla \psi^*({\bf x},t))
\psi({\bf x},t)\right]\,,\nonumber\\
&=&\left(\frac{\hbar}{m}\right) {\rm Im}\,(\psi^*({\bf x},t) \nabla
\psi({\bf x},t))\,,
\label{pf}
\end{eqnarray}
where `Im' implies the imaginary part of some quantity. Most of the
elementary quantum mechanics books then proceeds to show that:
\begin{eqnarray}
\int d^3x\, {\bf j}({\bf x}, t) = \frac{\langle \hat{\bf p} \rangle}{m}\,,
\label{jp}
\end{eqnarray}
which is obtained from Eq.~(\ref{pf}) by integrating both sides of it
over the whole volume. From Eq.~(\ref{pf}) we immediately see that if
the solution of the time-independent Schr\"{o}dinger equation is real we
will have ${\bf j}({\bf x}, t)=0$ and consequently from
Eq.~(\ref{jp}), $\langle \hat{\bf p} \rangle = 0$. But this statement
is also coordinate dependent, which is rarely said in elementary
textbooks of quantum mechanics. Eq.~(\ref{pf}) evidently does not hold
in spherical polar coordinates. If we take Eq.~(\ref{sepvar}) as the
solution in a general isotropic central potential and use the general
form of $\nabla$ in spherical polar coordinates then it can be seen
that ${\bf j}_r(r,\theta,\phi, t) = 0$ for a real potential. But then
Eq.~(\ref{jp}) does not hold as here $\hat{p}_r$ is simply the radial
component of $\nabla$ and not as given in Eq.~(\ref{remom}), and we
know $\langle \frac{d}{dr} \rangle$ is not zero. The reason why
Eq.~(\ref{pf}) is not suitable in spherical polar coordinates is
related to the fact that in deriving Eq.~(\ref{pf}) one assumes that
the probability density of finding the quantum state within position
${\bf x}$ and ${\bf x} + d{\bf x}$ at time $t$ is $|\psi({\bf x},
t)|^2$. But this statement is only true in Cartesian coordinates, in
spherical polar coordinates the probability density of the system to
be within a region $r$ and $r+dr$, $\theta$ and $\theta + d\theta$,
$\phi$ and $\phi + d\phi$ is not $|\psi(r,\theta,\phi)|^2$ but
$|\psi(r,\theta,\phi)|^2 r^2 \sin\theta$ and consequently the steps
which follow leading to Eq.~(\ref{pf}) in Cartesian coordinates are
not valid in spherical polar coordinates. In general Eq.~(\ref{pf})
will not be valid in any curvilinear coordinate system.

The next section contains the actual calculations of the expectation
values of the momentum operator in various cases where we have bound
state solutions.  In all the relevant cases discussed in this article
it is seen that although $\langle \hat{p}_x \rangle = 0$ but $\langle
\hat{p}^2_x \rangle$ is not zero as it is related to the Hamiltonian
operator. In all the cases we must have,
\begin{eqnarray}
\langle (\hat{p}_x)^{s} \rangle = 0\,,\,\,s={\rm odd\,\,\,integer}\,.
\label{gencond}
\end{eqnarray}
The above equation can be guessed from the reality of the expectation
value of the momentum operator.
%%%%%%%%%%%%%%%%%%%%%%%%%%%%%%%%%%%%%%%%%%%%%%
\section{Momentum expectation values in various bound states}
\label{momexpv}
%%%%%%%%%%%%%%
In this section we will calculate the momentum expectation values in
various bound states with stiff or slowly varying potentials.
%%%%%%%%%%%%%
\subsection{Particle in one-dimensional stiff potential wells}
%%%%%%%%%
\subsubsection{Infinite square well potential}
%%%%%%%%%%%%%
In this case we consider a particle to be confined in region
$-\frac{L}{2}$ to $\frac{L}{2}$ along the $x$-axis where the potential
is specified by,
\begin{eqnarray}
V(\hat{x})&=& \infty\,,\,|x| \geq \frac{L}{2}\,,\nonumber\\
    &=& 0\,,\, |x| < \frac{L}{2}\,.
\label{infwell}
\end{eqnarray}
In this case the solution of the time-independent Schr\"{o}dinger
equation, Eq.~(\ref{tise}), satisfies the boundary condition,
\begin{eqnarray}
u\left(-\frac{L}{2}\right)=u\left(\frac{L}{2}\right)=0\,,
\label{bciw}
\end{eqnarray}
and as the potential has parity symmetry about $x=0$ we have two sets
of solutions, the odd solutions:
\begin{eqnarray}
u^{(o)}_n(x)=\sqrt{\frac{2}{L}}\sin\left(\frac{2n\pi x}{L}\right)\,,
\label{odd}
\end{eqnarray}
and the even solutions:
\begin{eqnarray}
u^{(e)}_n(x)=\sqrt{\frac{2}{L}}\cos\left(\frac{(2n-1)\pi x}{L}\right)\,.
\label{even}
\end{eqnarray}
In the above equations $n$ is a positive integer. Both of these
functions, $u^{(o)}_n(x)$ for the odd case and $u^{(e)}_n(x)$ for the
even case, are real and are not momentum eigenstates. But the momentum
expectation values can be found out from the above solutions. For the
odd solutions we have:
\begin{eqnarray}
\langle \hat{p}_x\rangle &=& -i\hbar\int^{\frac{L}{2}}_{-\frac{L}{2}}
u^{(o)}_n(x)\frac{d u^{(o)}_n(x)}{d x}\,dx\,,\nonumber\\
&=& -\frac{4in\pi\hbar}{L^2}\int^{\frac{L}{2}}_{-\frac{L}{2}}
\sin\left(\frac{2n\pi x}{L}\right)\cos\left(\frac{2n\pi x}{L}\right)\,,\nonumber\\
&=& 0\,,
\end{eqnarray}
as expected. Similarly for the even solutions it is also easy to
show that the expectation value of the momentum operator vanishes.
%%%%%%%%%%%%%%%%%%%%%%%%%%%%%%%%%%%%%%%%%%%%%%%%%%%%%%%%%%%
\subsubsection{Finite square well potential}
%%%%%%%%%%%%%%%%%
In this case,
\begin{eqnarray}
V(\hat{x})&=& 0\,,\,\,|x| \geq a\,,\nonumber\\
    &=& -V_0\,, |x| < a\,,\,\,(V_0 > 0)\,.
\end{eqnarray}
If we are not interested in the normalization constant of the bound
state solution then the solution of the time-independent Schr\"{o}dinger
equation in this case is:
\begin{eqnarray}
u(x) &\sim & e^{-\kappa |x|}\,,\,\,|x| > a\,,\nonumber\\
     &\sim & \cos(kx)\,,\,\,|x| < a\,,\,\,({\rm even\,\,parity})\nonumber\\
     &\sim & \sin(kx)\,,\,\,|x| < a\,,\,\,({\rm odd\,\,parity})\,,
\end{eqnarray}
where,
\begin{eqnarray}
k^2 &=& \frac{2m(-|E| + V_0)}{\hbar^2}\,,\\
\kappa^2 &=& \frac{2m|E|}{\hbar^2}\,.
\end{eqnarray}
In this case the expectation value of the momentum operator is:
\begin{eqnarray}
\langle \hat{p}_x \rangle &\sim & -i\hbar\int^{+\infty}_{-\infty}
u(x)\frac{d u(x)}{d x}\,dx\,,\nonumber\\
&\sim & -i\hbar\left[\kappa \left(\int^{-a}_{-\infty} e^{2\kappa x} dx
 - \int^{\infty}_{a} e^{-2\kappa x}dx \right)
+ k\int^{a}_{-a} \sin(kx)\cos(kx)\,dx \right]\,,\nonumber\\
&=& 0\,,
\end{eqnarray}
where the first two lines of the above equation holds up to a constant
arising from the normalization of the wave-function. In deriving the
last equation we have taken the odd parity solution, but the result
remains unaffected if we take the even parity solution as well.
%%%%%%%%%%%%%%%%%%%%%%%%%%%%%%%%%%%%%%%%%%%%%%%%%%%%%%%%%%%%
\subsubsection{Dirac-delta potential}
%%%%%%%%%%%%%
In this case the potential is:
\begin{eqnarray}
V(\hat{x}) = -V_0\,\delta(\hat{x})\,,\,\,(V_0 > 0)\,.
\label{ddpot}
\end{eqnarray}
In this case there can be one bound state solution which is
obtained after solving the Eq.~(\ref{tise}). Demanding that the
solution $u(x)$ satisfies the boundary conditions:
\begin{eqnarray}
u(x=-\epsilon)&=&u(x=+\epsilon)\,,\\
\left.\frac{d u}{dx}\right|_{x=+\epsilon} - \,\,\,\,\,\,\,\left.\frac{d u}{dx}\right|_{x=-\epsilon}&=&
-\frac{2mV_0}{\hbar^2}\,u(x=0)\,,
\end{eqnarray}
where $\epsilon$ is an infinitesimal quantity tending to zero, we get
the form of the solution which is:
\begin{eqnarray}
u(x)&=& \sqrt{\kappa}\,e^{\kappa x}\,,\,\, x\leq 0\,,\\
    &=& \sqrt{\kappa}\,e^{\kappa x}\,,\,\, x\geq 0\,,
\label{ddsol}
\end{eqnarray}
where $\kappa=\frac{mV_0}{\hbar^2}$ and the energy of the bound state is
$E=-\frac{mV^2_0}{2\hbar^2}$.

The expectation value of the momentum operator in this case is:
\begin{eqnarray}
\langle \hat{p}_x \rangle &=& -i\hbar \int^{\infty}_{-\infty}
u(x)\frac{d u(x)}{d x}\,dx\,,\nonumber\\
&=&-i\hbar\kappa \left[\int^0_{-\infty} e^{2\kappa x}\,dx
- \int^{\infty}_0 e^{-2\kappa x}\,dx\right]\,,\nonumber\\
&=& 0\,.
\end{eqnarray}
In this case, also from Hermiticity of the momentum operator we
see that Eq.~(\ref{gencond}) holds true.
%%%%%%%%%%%%%%%%%%%%%%%%%%%%%%%%%%%%%%%%%%%%%%%%%%
\subsection{Particle in one-dimensional slowly varying potentials}
%%%%%%%%%%%
\subsubsection{Linear harmonic oscillator potential}
%%%%%%%%%%%%
In the case of the linear harmonic oscillator we have:
\begin{eqnarray}
V(\hat{x})=\frac12 m \omega^2 \hat{x}^2\,,
\end{eqnarray}
where $\omega$ is the angular frequency of the oscillator. The
solution of Eq.~(\ref{tise}) in this case, using the series solution
method, yields:
\begin{eqnarray}
u_n(q)=N_n\,e^{-\frac{q^2}{2}} H_n(q)\,,
\label{lhos}
\end{eqnarray}
where $n=0,1,2,\cdot,\cdot$ and $q=\sqrt{\alpha} x$ where
$\alpha=\frac{m\omega}{\hbar^2}$. $H_n(q)$ are Hermite
polynomials of order $n$ and $N_n$ is the normalization constant given
by,
\begin{eqnarray}
N_n=\left(\frac{1}{\sqrt{\pi}\,n!\,2^n}\right)^{\frac12}\,.
\end{eqnarray}
The momentum expectation value in this case turns out to be,
\begin{eqnarray}
\langle \hat{p}_x \rangle &=& -i\hbar\sqrt{\alpha}
\int^{\infty}_{-\infty}u_n(q)\frac{d u_n(q)}{d q}\,dq\,,\nonumber\\
&=&-i\hbar\sqrt{\alpha}N^2_n
\left[\int^{\infty}_{-\infty} e^{-q^2}H_n(q)\frac{d H_n(q)}{d q}\,dq
- \int^{\infty}_{-\infty} q\,e^{-q^2} H^2_n(q)\,dq \right]\,,\nonumber\\
&=& 0\,.
\end{eqnarray}
The first integral on the right side of the second line of the last
equation vanishes because, $\frac{d H_n(q)}{d q}=2n H_{n-1}(q)$ and
consequently the integral transforms into the orthogonality condition
of the Hermite polynomials. The second integral on the second line of
the right side of the above equation vanishes because the integrand is
an odd function of $q$.

The linear harmonic oscillator (LHO) has some very interesting
properties. To unravel them we have to digress a bit from the
wave-mechanics approach which we have been following and follow the
Dirac notation of {\it bra} and {\it kets}. The Hamiltonian of the LHO in
one-dimension is:
\begin{eqnarray}
\hat{H} = \frac{\hat{p}^2_x}{2m} + \frac12 m\omega^2 \hat{x}^2\,,
\end{eqnarray}
which can also be written as:
\begin{eqnarray}
\hat{H} = \hbar \omega \left( \hat{a}^\dagger \hat{a} + \frac12 \right)\,,
\end{eqnarray}
where $\hat{a}$ and $\hat{a}^\dagger$ are the annihilation and the creation
operators given by:
\begin{eqnarray}
\hat{a}\equiv \sqrt{\frac{m\omega}{2\hbar}}\left(\hat{x} +
\frac{i\hat{p}_x}{m\omega}\right)\,,\,\,\hat{a}^\dagger\equiv
\sqrt{\frac{m\omega}{2\hbar}}\left(\hat{x} - \frac{i\hat{p}_x}{m\omega}
\right)\,.
\label{aadag}
\end{eqnarray}
It can be seen clearly from the above definitions that $\hat{a}$ is
not an Hermitian operator. More over from the definition of the
operators we see that,
\begin{eqnarray}
[\hat{a}, \hat{a}^\dagger] = 1\,.
\end{eqnarray}
Conventionally the number operator is defined as:
\begin{eqnarray}
\hat{N}\equiv \hat{a}^\dagger \hat{a}\,,
\end{eqnarray}
and its eigen-basis are the number states $|n\rangle$ such that,
\begin{eqnarray}
\hat{N}\,|n\rangle = n |n\rangle\,.
\end{eqnarray}
The Hamiltonian of the LHO can be written in terms of the number
operator and consequently the number states are energy eigenstates.
In this basis the action of the annihilation and creation operators
are as:
\begin{eqnarray}
\hat{a}|n\rangle &=& \sqrt{n}\,|n-1\rangle\,,
\label{ann}\\
\hat{a}^\dagger |n\rangle &=& \sqrt{n+1}\,|n + 1\rangle\,.
\label{crea}
\end{eqnarray}
From the definitions of the annihilation and creation operators we can
write the momentum operator as:
\begin{eqnarray}
\hat{p}_x = -i\sqrt{\frac{m\hbar \omega}{2}}(\hat{a} - \hat{a}^\dagger)\,.
\label{lhom}
\end{eqnarray}
From Eq.~(\ref{ann}), Eq.~(\ref{crea}) and the above equation we can
write the matrix elements of the momentum operator as:
\begin{eqnarray}
\langle n'|\hat{p}_x |n\rangle = i\sqrt{\frac{m\hbar \omega}{2}}
\left(-\sqrt{n}\,\delta_{n',\,n-1} + \sqrt{n+1}\,\delta_{n',\,n+1}\right)\,.
\label{matmom}
\end{eqnarray}
The above equation shows that the momentum operator can connect two different
energy eigenstates.

In the case of LHO, except the number operator states, we can have
another state which is an eigenstate of the annihilation operator
$\hat{a}$. This state is conventionally called the coherent state and
it is given as:
\begin{eqnarray}
|\alpha \rangle = e^{-\frac{|\alpha|^2}{2}} \sum^\infty_{n=0}
\frac{\alpha^n}{\sqrt{n!}}|n\rangle\,,
\label{cohs}
\end{eqnarray}
where $\alpha$ is an arbitrary complex number. Now from
Eq.~(\ref{lhom}) we can find the momentum expectation value of the
coherent state and it is,
\begin{eqnarray}
\langle \hat{p}_x \rangle_\alpha \equiv \langle \alpha |\hat{p}_x |\alpha
\rangle = \sqrt{\frac{m\hbar \omega}{2}}\,\, \rm{Im}(\alpha)\,.
\label{cohm}
\end{eqnarray}
From the above equation we can see that although the expectation value
of the momentum operator is zero in the energy eigen-basis but it is
not so when we compute the momentum expectation value in the coherent
state basis, which is essentially a superposition of energy
eigenstates. It must be noted that the momentum expectation value is
non zero only when the parameter $\alpha$ has an imaginary part.
%%%%%%%%%%%%%%%%%%%%%%%%%%%%%%%%%%%%%%%%%%%%%%%%%%%%%%%%%%%%
\subsubsection{P\"{o}schl-Teller potential}
%%%%%%%%%%%
Among the potentials belonging to the hypergeometric class the P\"{o}schl-Teller potentials have been the most extensively studied and used. This class of potentials consist of trigonometric as well as the hyperbolic type. The trigonometric versions have found
applications in molecular and solid state physics and the hyperbolic
variants have been used in various studies related to black hole perturbations.

In the present work we use the trigonometric, symmetric
P\"{o}schl-Teller potential given by:
\begin{equation}
V(\hat{x}) = V_0 \tan^2 (a\hat{x})\,,
\end{equation}
where $V_0$ can be parameterized as:
\begin{equation}
V_0 = \frac{\hbar^2 a^2}{2m} \lambda (\lambda -1)\,,
\end{equation}
with for some positive number $\lambda > 1$ and $a$ is some scaling
factor. The energy eigenvalues of the bound state solutions are:
\begin{equation}
E_n = -\frac{\hbar^2 a^2}{2m} (n^2 + 2 n \lambda + \lambda)\,,
\end{equation}
and the solution of the time-independent Schr\"{o}dinger equation is,
\begin{equation}
u_n(x) = N_n \sqrt{\cos (a x)}
\, P_{n + \lambda - 1/2}^{1/2 - \lambda} \left(\sin (a x)\right)\,,
\label{pts}
\end{equation}
where,
\begin{equation}
N_n = \left[\frac{a (n+\lambda)
\Gamma(n+2\lambda)}{\Gamma(n+1)} \right]^{1/2}\,,
\end{equation}
is the normalization constant and $P_\nu^\mu(x)$ is the associated
Legendre function. At this point it is fair to point out that
$P^\mu_\nu(x)$ is not the Legendre polynomial $P^L_M(x)$ appearing in
Eq.~(\ref{thetaeqn}), as $\mu$ and $\nu$ need not be integers as $L$
and $M$. $P_\nu^\mu(x)$ is not a polynomial but the function appearing
in the right hand side of Eq.~(\ref{pts}) is a polynomial.

Now as claimed in the text let us show that the momentum expectation
value is indeed zero. Before we proceed let us simplify the notation a
bit by calling $\mu=1/2 - \lambda$ and $\nu=n + \lambda -
1/2$. Substituting $z=ax$ we can write the momentum expectation value
as:
\begin{eqnarray}
\langle \hat{p}_x \rangle = -i \hbar N_n^2 \int^{\pi/2}_{-\pi/2} dz \,
\sqrt{\cos (z)} \, P_\nu^\mu \left(\sin (z) \right) \frac{d}{dz} \left(\sqrt{\cos
(z)} \, P_\nu^\mu \left(\sin (z)\right)\right)\,.
\end{eqnarray}
Note the limits of the integration range from $\pi/2$ to $- \pi/2$
since at this value the potential becomes infinity hence we need not
consider the integration range to be the whole real line. For the sake
of convenience let us make a change of variable; letting $y = \sin(z)$
the above integral becomes:
\begin{equation}
\langle \hat{p}_x \rangle = -i \hbar N_n^2 \int^{+1}_{-1} dy \, (1 -
y^2)^{1/4} P_\nu^\mu (y) \frac{d}{dy} \left[(1 - y^2)^{1/4} P_\nu^\mu
(y)\right]\,.
\end{equation}
Taking the derivative inside the integral we get:
\begin{equation}
\langle \hat{p}_x \rangle = -i \hbar N_n^2 \int^{+1}_{-1} dy \left[(1 -
y^2)^{1/2} P_\nu^\mu (y) \frac{d P_\nu^\mu (y)}{dy} - \frac{y(1 -
y^2)^{-1/2}}{2} P_\nu^\mu (y) P_\nu^\mu (y) \right]\,.
\label{ptp}
\end{equation}
It is known that for associated Legendre functions \cite{grads1},
\begin{eqnarray}
P^\mu_\nu(-x) =\cos[(\mu + \nu)\pi]\,P^\mu_\nu (x) - \frac{2}{\pi}
\sin[(\mu + \nu)\pi]\,Q^\mu_\nu (x)\,,
\label{lfp}
\end{eqnarray}
where $Q^\mu_\nu(x)$ is the other linearly independent solution of
the associated Legendre differential equation. As in our case $\mu
+ \nu = n$ so $P^\mu_\nu(x)$ will have definite parity.  As
$P_\nu^\mu (x)$ has definite parity so the contribution of the
second term in the above integral vanishes since the total
integrand is an odd function.  The first integral is similar to
the one in Eq.~(\ref{postel}) and, due to the typical parity
property of $P^\mu_\nu(x)$ as shown in Eq.~(\ref{lfp}), it also
vanishes. Consequently we have $\langle \hat{p}_x \rangle = 0 $ as
expected.
%%%%%%%%%%%%%%%%%%%%%%%%%%%%%%%%%%%%%%%%%%%%%%%%%%%%%%%
\subsubsection{Morse potential}
\label{morse}
%%%%%%%%%%%%%%%%%%%%%%%%%%%%%%%%
Diatomic molecule is an exactly solvable system, if one neglects
the molecular rotation. The most convenient model to describe the
system, is the Morse potential \cite{Morse}:
\begin{equation}
V (\hat{x})=D(e^{-2\beta \hat{x}}-2e^{-\beta \hat{x}})\,,
\label{potential}
\end{equation}
where $x=r/r_{0}-1$, which is the distance from the equilibrium
position scaled by the equilibrium value of the inter-nuclear
distance $r_{0}$. $D$ is the depth of the potential, called
dissociation energy of the molecule and $\beta$ being a parameter
which controls the width of the potential.

 In terms of the above scaled variable $x$, the time-independent Schr\"{o}dinger equation becomes:
\begin{equation}
-\frac{\hbar^2}{2 \mu r_{0}}\frac{d^2 u(x)}{dx^2}+D(e^{-2\beta
x}-2e^{-\beta x}) u(x)=E u(x)\,.
\end{equation}
Here $\mu$ is the reduced mass of the molecule and the
corresponding bound state eigen function comes out to be:
\begin{equation}
u_{n}^{\lambda} (\xi)= N e^{-\xi/2} \xi^{s/2} L_{n}^{s} (\xi)\,,
\label{eigenstate}
\end{equation}
where the variables are described as,
\begin{equation}
\xi=2\lambda e^{-y}; \;\;y=\beta x; \;\;0 < \xi< \infty\,,
\label{variable}
\end{equation}
and
\begin{equation}
n=0,1,...,[\lambda-1/2]\,,
\end{equation}
which is nothing but the quantum number of the vibrational bound
states. Here $[\rho]$ denotes the largest integer smaller than
$\rho$, thus total number of bound states is $[\lambda-1/2]+1$.
The parameters,
\begin{equation}
\lambda=\sqrt{\frac{2\mu D
r^{2}_{0}}{\beta^2\hbar^2}}\;\;\mathrm{and}\;\; s=\sqrt{-\frac{8\mu
r^{2}_{0}}{\beta^2\hbar^2} {\it E}}\,,
\end{equation}
satisfy the constraint condition $s+2n=2\lambda-1$. We note that the
parameter $\lambda$ is potential dependent and $s$ is related to
energy $E$.  In Eq.~(\ref{eigenstate}), $L_{n}^{s}(y)$ is the
associated Laguerre polynomial and $N$ is the normalization constant
\cite{Suranjana}:
\begin{equation}
N=\left[\frac{\beta(2\lambda-2n-1)\Gamma{(n+1)}}{\Gamma{(2\lambda-n)r_0}}
\right]^{1/2}\,.
\label{mnorm}
\end{equation}

We are looking for the expectation value of linear
momentum for a vibrating diatomic molecule, and its expression is:
\begin{equation}
\langle \hat{p}_x \rangle=-i\hbar \int^{\infty}_{-\infty} u_{n}^{*}
(\xi)\frac{d}{dx} u_{n} (\xi)dx\,.
\end{equation}
In terms of the changed variable $\xi=2\lambda e^{-\beta x}$ the
integration limit changes to $\infty$ to $0$ and the expectation
value becomes:
\begin{eqnarray}
\langle \hat{p}_x \rangle &=&-i\hbar \int^{0}_{\infty} u_{n}^{*}
(\xi)\frac{d}{d\xi} u_{n} (\xi)d\xi\nonumber\\
&=& i\hbar N^2
\left[ -\frac{1}{2} \int_{0}^{\infty}e^{-\xi} \xi^{s} (L_{n}^{s}
(\xi))^2 d\xi+\frac{s}{2}\int_{0}^{\infty}e^{-\xi} \xi^{s-1}
(L_{n}^{s}(\xi))^2 d\xi\right.\nonumber\\
&+&\left.\int_{0}^{\infty}e^{-\xi} \xi^{s}
L_{n}^{s}(\xi)\frac{d}{d\xi}L_{n}^{s} (\xi)d\xi\right]\nonumber\\
&=& i\hbar N^2\left[-\frac{1}{2}I_1 +\frac{s}{2} I_2 + I_3
\right]\,.
\label{i1i2}
\end{eqnarray}
Integral $I_1$ is the orthogonality relation of the associated
Laguerre polynomials, which is:
\begin{equation}
\int_{0}^{\infty}e^{-\xi} \xi^{s} L_{n}^{s} (\xi) L_{m}^{s} (\xi)
d\xi= \frac{\Gamma(s+n+1)}{\Gamma(n+1)}\,\delta_{m, n}\,.
\label{lortho}
\end{equation}
To evaluate the second integral one uses the normalization integral
of Morse eigenstates. The normalization relation is:
\begin{equation}
\int_{-\infty}^{\infty} u^*(\xi) u(\xi)dr=\frac{|N|^2
r_0}{\beta} \int_{0}^{\infty}e^{-\xi} {\xi}^{s-1} (L_{n}^{s}
(\xi))^2 d\xi=1\,.
\end{equation}
The above integral involving $\xi$, is explicitly $I_2$.
$N$, being the normalization constant as given in
Eq.~\ref{mnorm}. Thus it is very straight forward to evaluate $I_2$
from the above relation as,
\begin{equation}
I_2=\frac{\Gamma(n+s+1)}{s\;\Gamma(n+1)}\,. \label{Itwo}
\end{equation}
The last integrand $I_3$ includes a differentiation which can be written as
\cite{Gradshteyn}:
\begin{equation}
\frac{d}{d\xi}L_{n}^{s}(\xi)=-L_{n-1}^{s+1}(\xi)\,.
\end{equation}
Writing the right hand side of the above equation as a summation
\cite{grads3}:
\begin{eqnarray}
L_{n}^{s+1}=\sum_{m=0}^{n}L_{m}^{s}\,,
\end{eqnarray}
and substituting the derivative term in integral $I_3$ we obtain:
\begin{equation}
I_3 = -\sum_{m=0}^{n-1}\int_{0}^{\infty}e^{-\xi} \xi^{s} L_{n}^{s}
(\xi) L_{m}^{s} (\xi) d\xi\,.
\end{equation}
In the above integral $m\neq n$ because $m$ can go only upto
$(n-1)$. Thus the integral vanishes. Now let us see what is the
expectation value of momentum observable, after evaluating the
three integrals above. Substituting the non-zero values $I_1$ and
$I_2$ in Eq.~\ref{i1i2}, it is clear that the expectation value of
momentum is zero as has been expected.

%%%%%%%%%%%%%%%%%%%%%%%%%%%%%%%%%%%%%%%%%%%%%%%%%%%%%%%
\subsection{Position expectation values for various potentials}
\label{expectx}
%%%%%%%%%%%%%%%%%%%%%%%%%%%%%%%%%%%%%%%%%%%%%%%%%%%%%%%
After a thorough discussion about the momentum expectation values for
various solvable one-dimensional potentials, it is worth spending some
time discussing about the average position of the particle
inside the bound states. Among all the above examples, in each case we
had $V(x)=V(-x)$ except the Morse potential as Morse potential is not
an example of a symmetric potential: $V(x)\ne V(-x)$.

In deriving the expectation values of momentum for above symmetric
cases, we often considered that the integrals of odd functions
over the symmetric limits vanishes. This result does not hold true
for the asymmetric Morse potential. Already we have shown that the
momentum expectation value: $<p>=0$ for all the above potentials. When
it comes to the expectation values of position, one can easily see
that $<x>=0$ for symmetric potentials whose centers are at the
origin. On the other hand if this is not the case, suppose the
infinite square well is defined in the range $0\le x \le L$ also then
the expectation value of position does not vanish. It becomes $L/2$.
Thus, more accurately the average position of the particle is
dependent on the symmetry of the potential where as the average
momentum is solely guided by the reality of it's eigenvalues and
consequently it is zero always.

Below we will briefly discuss how the asymmetry of the potential
affects the expectation value of $x$ in the case of the Morse
potential. The expectation value of the position operator is:
\begin{equation}
\langle {\hat x}\rangle =\int_{-\infty}^{\infty}u_{n}^{\lambda*}(\xi)\,x\,u_{n}^{\lambda}(\xi)dx.
\end{equation}
The eigen function and the variables are respectively substituted
from Eq.~(\ref{eigenstate}) and Eq.~(\ref{variable}). We obtain
\begin{equation}
\langle{\hat x}\rangle=\frac{N^{2}}{\beta^{2}}
\left[\ln(2\lambda)\int_{0}^{\infty}e^{-\xi} {\xi}^{s-1}
(L_{n}^{s} (\xi))^2 d\xi+\int_{0}^{\infty}e^{-\xi} {\xi}^{s-1}
(L_{n}^{s} (\xi))^2 \ln(\xi)d\xi\right].
\end{equation}
The first integral is already been obtained in Eq.~(\ref{Itwo}).  This
result is independent of the quantum number $n$. The second integral
(say $I$) is not that straight forward, because it contains associated
Laguerre polynomial, logarithm, exponential and monomial
functions. Here at best we can evaluate the integral atleast for some
specific $n$ as, $n=0$ or $n=1$, when the Laguerre polynomial is
respectively replaced by $1$ and $(-\xi+s+1)$. For the ground state
wave function ($n=0$), $I$ would be
\begin{equation}
I_{n=0}=\int_{0}^{\infty}e^{-\xi} {\xi}^{s-1}\ln(\xi)d\xi,
\end{equation}
which can be written in terms of $\Psi(s)$ and $\Gamma$ function \cite{grads4}:
\begin{equation}
I_{n=0}=\Gamma(s) \Psi(s),
\end{equation}
where, $\Psi(s)$ is the logarithmic factorial function, defined as
$\frac{d(\ln(s)!}{ds}=\frac{(s!)^{'}}{s!}=\Psi(s)$. For $n=0$,
first integral reduces to $\Gamma(s)$ from Eq.~(\ref{Itwo}). Above
two evaluations gives the ground state expectation value:
\begin{equation}
\langle \hat{x}\rangle_{n=0}=\frac{1}{r_{0}\beta}\left[\ln(s+1)-\Psi(s)\right].
\end{equation}
For $n=1$, one can proceed in the same way
\begin{eqnarray}
\langle{\hat x}\rangle_{n = 1} &=&\frac{N^{2}}{\beta^{2}}
\left[\int_{0}^{\infty}e^{-\xi} {\xi}^{s+1}\ln(\xi)d\xi + (s+1)^2
\int_{0}^{\infty}e^{-\xi} {\xi}^{s-1}\ln(\xi)d\xi
2(s+1)\int_{0}^{\infty}e^{-\xi}
{\xi}^{s}\ln(\xi)d\xi\right]\\\nonumber &=&\frac{N^{2}}{\beta^{2}}
\left[\Gamma(s+2) \Psi(s+2) + (s+1)^2 \Gamma(s) \Psi(s) -
2(s+1)\Gamma(s+1) \Psi(s+1)\right],
\end{eqnarray}
which simplifies to give the expectation value corresponding to
the second eigen state:
\begin{equation}
\langle{\hat x}\rangle_{n=1}=\frac{1}{r_{0}\beta}\left[\ln(s+3)
-\Psi(s+2)+\frac{3}{(s+2)})\right].
\end{equation}
Other expectation values for $n>1$ can also be obtained in a
similar fashion.

The important point which is to be noted here is,
though the average momentum vanishes, the average position is
non-zero for Morse potential and remain so, irrespective of the
choice of coordinate origin. This result is  also true for all
eigen states of the same Hamiltonian.
%%%%%%%%%%%%%%%%%%%%%%%%%%%%%%%%%%%%%%%%%%%%%%%%%%%%%%%
\subsection{Momentum expectation value for a three-dimensional slowly varying
spherically symmetric  potential} \label{hydr}
%%%%%%%%%%%%%%%%%%
In three dimensions, for a spherically symmetric potential the
solution of the Schr\"{o}dinger equation is given in
Eq.~(\ref{sepvar}). Here we have assumed that the variables can be
separated. The expectation values of $\hat{L}_\theta$ and
$\hat{L}_\phi$ have been evaluated in section \ref{intri}. In this
section we take the case of the Hydrogen atom and calculate the
expectation value of the radial component of the linear momentum.
%%%%%%%%%%%%%%%%%%%%%%%
\subsubsection{The Hydrogen atom}
%%%%%%%%%%%%%%%%%%
In this case,
\begin{eqnarray}
V(\hat{r})=-\frac{e^2}{\hat{r}}\,.
\end{eqnarray}
where $e$ is the electronic charge and $r=\sqrt{x^2 + y^2 + z^2}$. Now
we have to write Eq.~(\ref{tise}) in spherical polar coordinates and
the solution of the time-independent Schr\"{o}dinger equation is:
\begin{eqnarray}
u_{n \, L \, M}(r,\theta,\phi)
&= & N_r \, R_{n \, L}(r) \, Y_{L \, M}(\theta,\phi)\,,\nonumber\\
& = & N_r \, e^{-r/n a_0}
\left[\frac{2r}{n a_0} \right]^{L} {\cal L}_{n - L -1}^{2 L + 1}
\left(\frac{2r}{n a_0} \right) Y_{L \, M}(\theta,\phi)\,,
\label{hatsol}
\end{eqnarray}
where $a_0=\frac{\hbar^2}{m e^2}$ is the Bohr radius and $m$ is the
reduced mass of the system comprising of the proton and the electron.
$n$ is the principal quantum number which is a positive integer,
${\cal L}_{n - L -1}^{2 L + 1}(x)$ are the associated Laguerre
polynomials, $Y_{L \, M}(\theta,\phi)$ are the spherical-harmonics,
and $N_r$ is the normalization arising from the radial part of the
eigenfunction.  The values which $L$ and $M$ can take is discussed in
section \ref{intri}.  The radial normalization constant is given by:
\begin{equation}
N_r = \left[ \left(\frac{2}{n a_0} \right)^3 \frac{(n - L -
1)!}{(n + L)! 2n} \right]^{1/2}.
\end{equation}
The spherical-harmonics are given by,
\begin{equation}
Y_{L \, M}(\theta,\phi) = \left[\frac{2 \, L + 1}{4 \pi}
\frac{(L - M)!}{(L + M)!} \right]^{1/2} P^L_M(\cos \theta)
e^{i M \phi}\,,
\end{equation}
where $P_L^M(\cos \theta)$ are the associated Legendre functions. It
is noted that although the Coulomb potential is a real potential but
the solution in spherical polar coordinates is not real, $e^{i M
\phi}$, is complex.  The spherical-harmonics are ortho-normalized
according to the relation,
\begin{equation}
\int_{\theta=0}^\pi \int_{\phi=0}^{2 \pi} d \theta \, d \phi \, \sin
\theta \, Y_{L \, M}(\theta, \phi) Y_{\tilde{L}\,\tilde{M}}(\theta, \phi) \,
 = \delta_{L \tilde{L}} \, \delta_{M \tilde{M}}.
\end{equation}
Let us write the eigenfunctions in terms of dimensionless quantity:
$\rho = 2r/n a_0 \equiv \alpha r$. Also we define $k \equiv (2 L +
1)$ and $n_r \equiv (n - L - 1)$ for the sake of convenience.  With
this amount of notational machinery the eigenfunctions can be written
as:
\begin{equation}
u_{n \, L \, M}(r,\theta,\phi)  =  N_r \, R_{n L}(\rho)
\, Y_{L \, M}(\theta,\phi)\,.
\end{equation}
The radial momentum expectation value in this case is not given by
$-i\hbar\langle \frac{\partial}{\partial \rho}\rangle$, its form
is (already discussed in section \ref{intri}):
\begin{equation}
\langle \hat{p}_\rho \rangle = - i \hbar \tilde{N}^2 \int_0^\infty
d\rho \, \rho^2 R^\ast_{n L}(\rho) \left[\frac{\partial}{\partial
\rho} + \frac{1}{\rho} \right] R_{n L}(\rho) \int d\Omega \,
[Y_{L \, M}(\theta,\phi)]^2\,.
\end{equation}
Where $\tilde{N}^2 = N_r^2/\alpha^2$. The integral for the spherical
harmonics yields identity. The radial expectation value then becomes,
\begin{equation}
\langle \hat{p}_\rho \rangle = - i \hbar \tilde{N}^2 \int_0^\infty
d\rho \left\{- \frac{1}{2} e^{-\rho} \rho^{k+1} [\mathcal{L}_{n_r}^k(\rho)]^2
+ (L + 1) \, e^{-\rho} \rho^{k} [\mathcal{L}_{n_r}^k(\rho)]^2 + e^{-\rho} \rho^{k+1} \mathcal{L}_{n_r}^k(\rho) \frac{d}{d
\rho}[\mathcal{L}_{n_r}^k(\rho)] \right\}\,.
\end{equation}
Using the recurrence relation \cite{Gradshteyn}:
\begin{equation}
\frac{d}{d\rho} \mathcal{L}_{n_r}^k(\rho) = \rho^{-1} \left[n_r \, \mathcal{L}_{n_r}^k(\rho) - (n_r + k) \, \mathcal{L}_{n_r - 1}^k(\rho) \right],
\end{equation}
the expectation value integral acquires the form:
\begin{equation}
\langle \hat{p}_\rho \rangle = - i \hbar \tilde{N}^2 \int_0^\infty
d\rho \left\{- \frac{1}{2} e^{-\rho} \rho^{k+1} [\mathcal{L}_{n_r}^k(\rho)]^2
+ (n_r + L + 1) \, e^{-\rho} \rho^{k} [\mathcal{L}_{n_r}^k(\rho)]^2 + e^{-\rho} \rho^{k} \mathcal{L}_{n_r}^k(\rho) \mathcal{L}_{n_r -1}^k(\rho)
\right\}\,.
\end{equation}
The third contribution of the becomes zero from the orthogonality
property of the associated Laguerre polynomials as given in
Eq.~(\ref{lortho}).  The contribution from the second term can also be
found similarly.  To find the share of the first term we make use
of \cite{math}:
\begin{equation}
\int_0^\infty d\rho \, e^{-\rho} \, \rho^{k+1} \, [\mathcal{L}_{n_r}^k (\rho)]^2  =
\frac{(n_r + k)!}{n_r !} \, (2n_r + k +1).
\end{equation}
Collecting all the contributions we get the radial expectation value
to be zero as expected.
%%%%%%%%%%%%%%%%%%%%%%%%%%%%%%%%%%%%%%%%%%%%%%%%%%%%%%%%%%%%
\section{A discussion on Heisenberg's equation of motion and
Ehrenfest theorem}
\label{ether}
%%%%%%%%
The time evolution of any operator $\hat{O}$ in the Heisenberg picture
is given by:
\begin{eqnarray}
\frac{d \hat{O}}{dt}=\frac{1}{i\hbar}[\hat{O}, \hat{H}]\,,
\label{heqn}
\end{eqnarray}
where $\hat{H}$ is the Hamiltonian of the system. The Hamiltonian of a
quantum system comprising of a particle of mass $m$ is given by:
\begin{eqnarray}
\hat{H}=\frac{\hat{\bf p}^2}{2m} + V(\hat{\bf x})\,.
\label{ham}
\end{eqnarray}
From the above two equations we can write the time evolution of the
momentum operator in one dimension, in Cartesian coordinates as:
\begin{eqnarray}
\frac{d \hat{p}_x}{dt}=\frac{1}{i\hbar}[\hat{p}_x, \hat{H}]
= -\frac{d}{d x} V(\hat{x})\,,
\label{nlaw}
\end{eqnarray}
which is the operator version of Newton's second law in a time
independent potential. Now if we take the expectation values of both
sides of Eq.~(\ref{nlaw}) in any basis we get:
\begin{eqnarray}
\frac{d \langle \hat{p}_x \rangle }{dt}
= -\left\langle \frac{d}{d x} V(\hat{x}) \right\rangle\,,
\label{eht}
\end{eqnarray}
and historically the above equation is called the Ehrenfest theorem,
which was deduced in a different way by P. Ehrenfest. Using the
Ehrenfest theorem we can deduce that the rate of change of the
expectation value of the momentum operator is zero in the case of the
linear harmonic oscillator. In the case of the linear harmonic
oscillator we have:
\begin{eqnarray}
\frac{d}{d x} V(\hat{x})= m\omega^2 \hat{x}\,,
\end{eqnarray}
and it can be trivially shown that $\langle \hat{x} \rangle = 0$. This
directly implies that,
\begin{eqnarray}
\frac{d \langle \hat{p}_x \rangle }{dt} = 0\,,
\end{eqnarray}
for the linear harmonic oscillator. The above equation shows that the
expectation value of the momentum along $x$ direction is constant, and
this constant is zero is known from other sources.

Next we focus on the Hydrogen atom. The Hamiltonian of the Hydrogen atom is:
\begin{eqnarray}
\hat{H}= -\frac{\hbar^2}{2 m} \frac{1}{r}\frac{\partial^2}{\partial r^2}\,r
+ \frac{1}{2 m r^2} \hat{\bf L}^2 - \frac{e^2}{r}\,,
\label{hham}
\end{eqnarray}
where,
\begin{eqnarray}
\hat{\bf L}^2 = -\hbar^2\left(\frac{1}{\sin\theta} \frac{\partial}{\partial
\theta}\sin\theta\frac{\partial}{\partial \theta} + \frac{1}{\sin^2\theta}
\frac{\partial^2}{\partial \phi^2}\right)\,,
\end{eqnarray}
whose eigenvalues are of the form $\hbar^2 L(L+1)$ in the basis $Y_{L
\, M}(\theta, \phi)$. In the expression of the Hamiltonian $m$ is the
reduced mass of the system comprising of the proton and electron. Next
we try to apply Heisenberg's equation to the radial momentum
operator. Noting that the first term of the Hamiltonian is nothing but
$\hat{p}^2_r$ the Heisenberg equation is:
\begin{eqnarray}
\frac{d \hat{p}_r}{d t} &=&
- \frac{\hat{\bf L}^2}{2 m}\left[\frac{1}{r}\frac{\partial}{\partial r}\,
r\,,\,\frac{1}{r^2}\right] + e^2\left[\frac{1}{r}\frac{\partial}{\partial r}\,r\,,\,\frac{1}{r}\right]\,,\nonumber\\
&=& \frac{\hat{\bf L}^2}{m r^3} - \frac{e^2}{r^2}\,.
\label{hheis}
\end{eqnarray}
The above equation is the operator form of Newton's second law in
spherical polar coordinates. Next we evaluate the expectation value of
both the sides of the above equation using the wave-functions given in
Eq.~(\ref{hatsol}). We know,
\begin{eqnarray}
\left\langle \frac{1}{r^2} \right\rangle &=& \frac{1}{n^3\, a^2_0 (L +
\frac12)}\,,\\
\left\langle \frac{1}{r^3} \right\rangle &=& \frac{1}{a^3_0 \, n^3 L (L +
\frac12) (L+1)}\,.
\end{eqnarray}
Using the above expectation values in Eq.~(\ref{hheis}) and noting
that $\langle \hat{\bf L}^2 \rangle = \hbar^2 L (L + 1)$ we see that
the time derivative of the expectation value of the radial momentum
operator of the Hydrogen atom vanishes.

The above analysis shows that the form of the Ehrenfest theorem as
given in Eq.~(\ref{eht}) is only valid in Cartesian coordinates. In
the case of the Hydrogen atom if we used Eq.~(\ref{eht}) we should
have never got the correct result.
%%%%%%%%%%%%%%%%%%%%%%%%%%%%%%%%%%%%%%%%%%%%%%%%%%%%%%%%%%%
\section{Conclusion}
%%%%%%%%%%%%%%%%%%%%%
In the present work we have emphasized on the reality of the
momentum expectation value and using the reality of the
expectation value as a bench mark we did find out the form of the
momentum operator in spherical polar coordinate system. We found
that most of the concepts which define the momentum operator in
Cartesian coordinates do not hold good in spherical polar
coordinates and in general in any other coordinate system. The
reason being that whenever we do an integration in curvilinear
coordinates the Jacobian of the coordinate transformation matrix
comes inside the picture and the Cartesian results start to falter
if we do not change the rules appropriately. The forms of the
momentum along the radial direction and the form of the angular
momentum operators are derived in section \ref{intri}. The status
of the angular variables was briefly discussed in the same
section.  We explicitly calculated the expectation values of the
momentum operator in various important cases and showed that the
expectation value of the momentum operator do really come out to
be zero as expected. Although the expectation value of the
momentum operator vanishes in most of the bound states, with a
real potential, the expectation value of the position is not
required to vanish. The expectation value of the position operator
is directly related with the parity property of the potential
which was briefly discussed in subsection \ref{expectx}.  At the
end we calculated the Heisenberg equation of motion for the radial
momentum operator for the Hydrogen atom and showed its formal
semblance with Newton's second law. It was also shown that if we
properly write the Heisenberg equation of motion in spherical
polar coordinates then Ehrenfest's theorem follows naturally.

In short we conclude by saying:
\begin{enumerate}
\item the forms of the various momentum operators, in most of the
coordinate systems, in quantum mechanics can be obtained by imposing
the condition of the reality of their eigenvalues. The form of the
probability conservation equation and Ehrenfest theorem must be
modified in curvilinear coordinates to yield meaningful results.
\item There are obvious problems in elevating the status of angular variables
to dynamical variables in quantum mechanics.
\item For compact variables, if the variable is periodic the expectation value of the angular momentum conjugate to it is non-zero. If the compact variable is not periodic then the angular momentum conjugate to it must vanish.
\item The momentum expectation values in cases of bound state motions vanish, whereas the position expectation values in those cases depends on the symmetry of the potential.
\end{enumerate}
%
%%%%%%%%%%%%%%%%%%%%%%%%%%%%%%%%%%%%%%%%%%%%%%%%%%%%%%%%%%%%
\begin{center}
{\bf Acknowledgements}
\end{center}
%%%%%%%%%%%%%%%
The authors thank Professors D. P. Dewangan, S. Rindani, J. Banerji,
P. Panigrahi and Ms. Suratna Das for stimulating discussions and constant
encouragements.
%%%%%%%%%%%%%%%%%%%%%%%%%%%%%%%%%%%%%%%%%%%%%%%%%%%%%%%%%%%%

%%%%%%%%%%%%%%%
\end{document}